\documentclass[epj]{svjour}
\usepackage{color,amsmath}
\usepackage{amssymb,bm}
\usepackage{graphicx,cite}

\begin{document}

\newcommand{\nwc}{\newcommand}
\nwc{\la}{\langle}
\nwc{\ra}{\rangle}
\nwc{\nn}{\nonumber}
\nwc{\Ra}{\Rightarrow}
\nwc{\wt}{\widetilde}
\nwc{\td}{\tilde}
\nwc{\lw}{\linewidth}
\nwc{\dg}{\dagger}
\nwc{\mL}{\mathcal{L}}

\nwc{\Tr}[1]{\underset{#1}{\mbox{Tr}}~}
\nwc{\av}[1]{\left< #1\right>}
\nwc{\pd}[2]{\frac{\partial #1}{\partial #2}}
\nwc{\ppd}[2]{\frac{\partial^2 #1}{\partial #2^2}}

\title{Exchange fluctuation theorems for a chain of interacting particles in presence of two heat baths}
\author{Sourabh Lahiri\inst{1}\and A. M. Jayannavar\inst{2}} 

\institute{Korea Institute for Advanced Study, 85 Hoegiro, Dongdaemun-gu, Seoul 130-722, Republic of Korea \and Institute of Physics, Sachivalaya Marg, Bhubaneswar 751005, India}

\date{\today}

\abstract{
The exchange fluctuation theorem for heat exchanged between two systems at different temperatures, when kept in direct contact, has been investigated by C. Jarzynski and D. K. W\'ojcik, in Phys. Rev. Lett. {\bf 92}, 230602 (2004). We extend this result to the case where two Langevin reservoirs at different temperatures are connected via a conductor made of interacting particles, and are subjected to an external drive or work source. The Langevin reservoirs are characterized by  Gaussian white noise fluctuations and concomitant friction coefficients. We first derive the Crooks theorem for the ratio between forward and reverse paths, and discuss the first law in this model. Then we derive the modified detailed fluctuation theorems (MDFT) for the heat exchanged at each end. These theorems differ from the usual form of the detailed fluctuation theorems (DFT) in literature, due the presence of an extra multiplicative factor. This factor quantifies the deviation of our MFDT from the DFT. Finally, we numerically study our model, with only two interacting particles for simplicity.
}


\titlerunning{Exchange fluctuation theorem for a chain of interacting particles}
\maketitle

\section{Introduction}

The fluctuation relations (FTs) have been one of the major developments in the field of nonequilibrium statistical mechanics in the last two decades \cite{han10,sei05,sei08,jar97,jar97a,eva02,har07,rit03,kur07,zon03,zon04,dha04,cro99,kur98,cro98}.

 They have not only helped us to understand the second law better, but have also lead to the generalization of  the response theory by providing a method to calculate nonlinear response coefficients \cite{gas07}. The Jarzynski equality and Crooks theorem are useful tools to determine free energies of systems undergoing nonequilibrium processes. Importantly, the relations remain valid irrespective of how far the system is driven away from equilibrium. 

Several developments have taken place since the pioneering works \cite{eva93,gal95,jar97}. The FTs have been generalized to quantum systems \cite{han11_rmp}. In quantum systems, it has been shown that although the FTs remain unaltered even when intermediate projective measurements of arbitrary observables are performed, the form of work distributions gets changed \cite{han11_pre}.
The FTs have also been generalized to systems that are driven by feedback-controlled forces \cite{sag12_pre,lah12a_jpa}. Under the action of feedback control, the modified theorems imply that the conventional form of the second law can be violated, with the discrepancy depending on the amount of information gained about the system, during the process. Steady state fluctuation theorems have also been proposed for heat engines, where the system is connected to two heat baths, and is being perturbed by an external periodic protocol \cite{sin11_jpa,lah12_jpa,cam14_arxiv}.

The steady state fluctuation theorem for heat has always attracted interest \cite{zon03,pug06,har06_epl,dha07_prl}, because it seems to hold for some model systems, but not in others. The form of heat fluctuation theorems that are studied in literature, in presence of a single heat bath, is given by
\begin{equation}
\lim_{\tau\to\infty}\frac{P(Q(\tau))}{P(-Q(\tau))} = e^{\beta Q(\tau)}.
\end{equation}
Here,  $\tau$ is the time of observation, whie $\beta$ is the inverse temperature of the bath in which the system is present. $P(Q(\tau))$ and $P(-Q(\tau))$ are the probability distributions for heat dissipated and absorbed during the process up to time $\tau$, respectively.

The fluctuation theorem for exchanged heat was studied in \cite{jar04}, where two bodies at different temperatures $T_1$ and $T_2$ were kept in contact, and the heat exchanged $Q_x$, follows the fluctuation theorem:
\begin{align}
\frac{P(Q_x)}{P(-Q_x)} = e^{\Delta\beta Q_x},
\end{align}
where $\Delta\beta = T_2^{-1}-T_1^{-1}$, where the Boltamann constant has been set to unity.
This is an exact result, and universal character of distributions depends only on the two temperatures and not on any system parameter. However, it may be noted that if we have initially prepared thermal conductors at temperatures $T_1$ and $T_2$, brought into contact without any intervening medium, then the heat lost by one system is naturally compensated by the heat gained by the other.  Gomez-Marin and Sancho \cite{gom06_pre} suggested that this theorem has to be modified when coupling mechanism between the two baths is considered. For this, they studied a specific model of ratchet, pawl and spring Brownian motor. Within this model, they also studied the case where load is present. 

In this work, we study the model consisting  of a chain of interacting particles, with the end particles being connected to two Langevin baths at different temperatures, and in the presence of time-dependent external drive. This is essentially a model of heat conduction from a hotter to a cooler heat bath, through a conductor \cite{dha08}. We will analyze this system in detail, and calculate the  fluctuation relations for the heat exchanged by the end particles with the corresponding heat baths. Although we deal with the one-dimensional case, this model can be readily generalized to higher dimensions. A more general case, where the system is connected to several baths with different temperatures and chemical potentials has been discussed in \cite{esp14_arxiv}, with the constraint that the initial state of the system should be sampled from an equilibrium distribution with respect to a reference reservoir.

\section{The model}

\begin{figure}[!h]
\centering
\includegraphics[width=\lw]{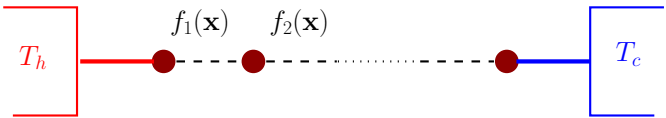}
\caption{Figure showing the heat conductor as a series of interacting particles, the end particles being connected to Langevin heat baths. The interaction forces need not be simply harmonic forces, and are in general denoted by $f_i(\bm x)$.}
\label{fig:mult}
\end{figure}

The system consists of a chain of $n$ particles of mass $m$, connected via interaction forces $f_i(\bm x)$ (see figure \ref{fig:mult}). Here, $\bm x$ denotes the coordinates of the particles to which the $i^{th}$ particle is interacting. Particle 1 and particle $n$ are connected to heat baths at temperatures $T_h$ and $T_c$, respectively, where $T_h>T_c$. In addition, an external perturbation, given by the onsite force term $g_i(x_i,t)$, also acts on the $i^{th}$ particle. The  equations of motion followed by the particles are as follows:
\begin{subequations}
\begin{align}
m\dot v_1 &= -\gamma_h v_1+f_1(\bm x)+g_1(x_1,t)+\xi_h(t); \label{eq:particle_1}\\
m\dot v_i &= f_i(\bm x)+g_i(x_i,t); ~~~\mbox{($i\ne 1,n$)};\label{eq:other_particles}\\
m\dot v_n &= -\gamma_c v_n+f_n(\bm x)+g_n(x_n,t)+\xi_c(t).\label{eq:particle_n}
\end{align}
\end{subequations}
Here, the particles at the extreme left (particle 1) and at the extreme right (particle 2) follow Langevin equations (eqs. \eqref{eq:particle_1} and \eqref{eq:particle_n}), being in direct contact with heat baths. The other particles, however, follow deterministic equations, given by eq. \eqref{eq:other_particles}. $\gamma_h$ and $\gamma_c$ are the friction coefficients associated with the hot and cold bath, respectively. $\xi_h(t)$ and $\xi_c(t)$ are the stochastic force terms that act on the particles directly connected to the two baths. They are assumed to be zero-mean Gaussian white noise: $\la\xi_a(t)\xi_b(t')\ra=2D_a\delta_{ab}\delta(t-t')$, and $\la\xi_a(t)\ra=0$, where the indices represent the bath labels: $a=h,c$ and $b=h,c$. $D_a=\gamma_aT$ is the noise strength of the heat bath $a$.
The force acting between the particle $i$ and its neighbours is given by 
\begin{align}
f_i(\bm x) = -\pd{H_1(\bm x)}{x_i}, 
\end{align}
where 
\begin{align}
H_1(\bm x) = \sum_{i=1}^{n-1}U_i(x_i-x_{i+1}).
\end{align}
$U_i(x_i-x_{i-1})$ is the interaction potential between particles $i$ and $i-1$.
The treatment goes through even when the interaction is not restricted to nearest neighbours.
The onsite force term $g_i(x_i,t)$ is considered to be conservative, i.e. it can be derived from a time-dependent potential $V_i(x_i,t)$:
\begin{align}
g(x_i,t) = -\pd{V_i(x_i,t)}{x_i}.
\label{eq:g}
\end{align}
$V_i(x_i,t)$ is the source for external work done on the system.

To derive the fluctuation theorems, we will need the expression for the Onsager-Machlup path integral for the system in phase space \cite{dha04_jpa,lah14_pla}. To do this, we first rewrite the Langevin equations as a single equation in matrix form:
\begin{align}
m\dot{\bm v} &= -\bm\gamma^\dg \bm v + \bm f + \bm g + \bm\xi(t).
\end{align}
Here,
\begin{align}
\bm v &=\left(\begin{array}{c} v_1\\ v_2\\ \vdots \\ v_n \end{array}\right); \hspace{0.5cm}
\bm\gamma = \left(\begin{array}{c} \gamma_h\\ 0\\ \vdots \\ 0\\ \gamma_c \end{array}\right);  \hspace{0.5cm}
\bm f = \left(\begin{array}{c} f_1(\bm x)\\ f_2(\bm x)\\ \vdots \\ f_n(\bm x) \end{array}\right); \hspace{0.5cm} \nn\\
\bm g &= \left(\begin{array}{c} g_1(x_1)\\ g_2(x_2)\\ \vdots \\ g_n(x_n) \end{array}\right); \hspace{0.5cm}
\bm\xi(t) = \left(\begin{array}{c} \xi_h(t)\\ 0\\ \vdots \\ 0\\\xi_c(t) \end{array}\right).
\end{align}
The Onsager-Machlup path probability  is then given by
\begin{align}
P_+ &= \mathcal N \exp\left[-\frac{1}{2}\bm\xi^\dg(t)\cdot \bm\sigma^{-1}\cdot \bm\xi(t) \right] \nn\\
&= \mathcal N \exp\left[-\frac{1}{4}\int_0^\tau dt~\left\{\frac{(m\dot v_1+\gamma v_1-f_1-g_1)^2}{D_h}\right.\right.\nn\\
&\left.\left.\hspace{2cm}+\frac{(m\dot v_n+\gamma v_n-f_n-g_n)^2}{D_c}\right\}\right], 
\label{eq:OM_functional}
\end{align}
where we have used,
\begin{align}
\bm\sigma^{-1} = \frac{\Delta t}{2} \left(\begin{array}{cccc} 1/D_h &0 &\cdots &0 \\
0&0&0&0\\
\vdots&0&\ddots&0\\
0&0&0& 1/D_c
\end{array}\right).
\end{align}
In \eqref{eq:OM_functional}, $\mathcal N$ is a normalization constant.
In a similar manner, we get the path probability for the reverse process:
\begin{align}
P_- &= \mathcal N \exp\left[-\frac{1}{4}\int_0^\tau dt~\left\{\frac{(m\dot v_1-\gamma v_1-f_1-g_1)^2}{D_h}\right.\right.\nn\\
&\left.\left.\hspace{2cm}+\frac{(m\dot v_n-\gamma v_n-f_n-g_n)^2}{D_c}\right\}\right]. \nn
\end{align}
On simplification, we obtain the following expression for the path ratio:
\begin{align}
\frac{P_+}{P_-} &= \exp\left[-\beta_h \int_0^\tau dt ~v_1(m\dot v_1-f_1-g_1) \right.\nn\\
&\left.\hspace{1cm}-\beta_c \int_0^\tau dt ~v_n(m\dot v_n-f_n-g_n)\right] \nn\\
&= e^{\beta_h Q_h + \beta_c Q_c},
\label{pathratio}
\end{align}
where $Q_h$ and $Q_c$ are the heats dissipated by the system into the hot and the cold baths, respectively. This expression comes from the stochastic definition of heat \cite{sek97,sek98}:
\begin{align}
Q_h &= \int_0^\tau dt ~v_1 (\gamma v_1-\xi_1(t)) \nn\\
&= \int_0^\tau dt ~v_1[f_1(\bm x)+g_1(x_1,t)-m\dot v_1],
\label{def:Q_h}
\end{align}
and similar definition holds for $Q_c$. In the last step, we have made use of the Langevin equations.
Here, all products follow the Stratonovich scheme, i.e., $v_1=[v_1(t)+v_1(t+\Delta t)]/2$ and $v_2=[v_2(t)+v_2(t+\Delta t)]/2$. Throughout this manuscript, we will use this discretization scheme, because the normal laws of calculus can then be applied. Eq. \eqref{pathratio} is the Crooks fluctuation theorem for the ratio of forward to the reverse path \cite{cro98,cro00}.

\section{The first law}

The energy balance equation can be directly obtained from the stochastic definitions for $Q_h$ and $Q_c$ (see eq. \eqref{def:Q_h}):
\begin{align}
-Q_h&-Q_c \nn\\
&= \sum_{i=1,n} \left[\frac{1}{2}m\{v_i^2(\tau)-v_i^2(0)\} - \int_0^\tau dt~v_if_i(\bm x)\right.\nn\\
&\hspace{4cm}\left.- \int_0^\tau dt~v_ig_i(x_i)\right] \nn\\
&= \sum_{i=1,n} \left[\frac{1}{2}m\{v_i^2(\tau)-v_i^2(0)\} - \int_0^\tau dt~v_if_i(\bm x)\right.\nn\\
&\hspace{4cm}\left.+ \int_0^\tau dt~v_i\pd{V_i}{x_i}\right] \nn\\
&= \sum_{i=1,n} \left[\frac{1}{2}m\{v_i^2(\tau)-v_i^2(0)\}+\Delta V_i \right.\nn\\
&\hspace{2cm}\left.- \int_0^\tau dt~v_if_i(\bm x)- \int_0^\tau dt\pd{V_i}{t}\right] \nn\\
&= \sum_{i=1,n} \left[\Delta E_i - W_i\right] \nn\\
&= \sum_{i=1}^n \left[\Delta E_i - W_i\right]\nn\\
\Ra Q_c &= -Q_h-\Delta E+W,
\end{align}
where we have used the relation $g_i(x_i,t)=-\partial V(x_i,t)/\partial x_i$. Here, $\Delta E$ denotes the change in the internal energy of the system, and $W$ is the thermodynamic work done \emph{on} the system.  $\Delta V_i \equiv V_i(x_i(\tau),\tau)-V_i(x_i(0),0)$ is the change in the potential acting on the $i^{th}$ particle due to external drive $g_i(t)$ (see eq. \eqref{eq:g}). We  have,
\begin{align}
\Delta E_i &= \frac{1}{2}m\{v_i^2(\tau)-v_i^2(0)\}+\Delta V_i- \int_0^\tau dt~v_if_i(\bm x);\nn\\
\Delta E &= \sum_{i=1}^n \Delta E_i; \hspace{0.5cm} W=\sum_{i=1}^n W_i; \hspace{0.5cm} W_i = \int_0^\tau dt \pd{V_i}{t}.\nn
\end{align}
Note that in the fifth step, we have converted the summation over 1 and $n$ to the summation over all particles, because the expression $\left[\Delta E_i  - W_i\right]$ becomes zero for all particles, except for the ones at the ends. This fact readily follows from the equations of motion for particle $i$ ($i\neq 1,n$):
\begin{align}
m\dot v_i &= f_i(\bm x)-\pd{V_i}{x_i} \nn\\
\Ra \int_0^\tau & dt~v_i\left[m\dot v_i - f_i(\bm x) + \pd{V_i}{x_i}\right]=0 \nn\\
\Ra \frac{1}{2}m&\{v_i^2(\tau)-v_i^2(0)\} - \int_0^\tau dt~v_if_i(\bm x) \nn\\
&\hspace{2cm}+\Delta V_i -W_i=0.
\end{align}

The first law in the above form implies that the energy gained in the form of work and absorbed heat, equals the increase in the total internal energy (including kinetic, potential and correlation energies) of the particles.

\section{The detailed fluctuation theorem}

By using the first law equality in the expression for path ratio \eqref{pathratio}, we obtain
\begin{align}
\frac{P_+}{P_-} &= e^{\beta_h Q_h - \beta_c(Q_h+\Delta E-W)} = e^{(\beta_h-\beta_c)Q_h-\beta_c(\Delta E-W)}.
\end{align}
Multiplying both sides by the ratio of initial probability distributions for the forward and the reverse processes, given by $p_0(\bm x_0,\bm v_0)$ and $p_1(\bm x_\tau,\bm v_\tau)$, we get
\begin{align}
\frac{p_0P_+}{p_1P_-} &= e^{(\beta_h-\beta_c)Q_h-\beta_c(\Delta E-W)+\Delta s}. 
\end{align}
Here, the boldfaced variables imply the full set of coordinates and velocities of all the particles. $\Delta s$ is the change in system entropy, given by $\Delta s = \ln (p_0/p_1)$. We have chosen the initial distribution $p_1(\bm x_\tau,\bm v_\tau)$ for the reverse process to be  the final distribution attained in the forward process. 

In the steady state at time $t$,  the steady state distribution is given by the form $p_s(\bm x,\bm v;t)=e^{-\phi(\bm x,\bm v;t)}$. Then, if the system begins and ends in steady states, we have $\Delta s(t)=\Delta \phi(t)$. Then the above path ratio can be used to derive the ratios of joint probabilities \cite{gar12} of $Q_h$, $\Delta E$, $W$ and $\Delta \phi$:
\begin{align}
&\frac{P_f(Q_h,\Delta E,W,\Delta \phi)}{P_r (-Q_h,-\Delta E,-W,-\Delta \phi)} \nn\\
&\hspace{0.5cm}= \exp\left[(\beta_h-\beta_c)Q_h+\beta_c(W-\Delta E)+\Delta \phi\right].
\label{central}
\end{align}
The subscripts $f$ and $r$ refer to the forward and reverse processes respectively.
Note that a similar relation has recently been derived for a quantum heat engine connected to two heat reservoirs at different temperatures \cite{cam14_arxiv} .
A direct consequence of eq. \eqref{central} is the relation
\begin{align}
\la e^{-(\beta_h-\beta_c)Q_h}\ra_f = \la e^{\beta_c(\Delta E-W)-\Delta \phi}\ra_r. 
\label{IFT_Qh}
\end{align}
The subscripts of the angular brackets imply the (forward or reverse) process along which the average has been computed.
$\Delta\phi$ is simply the change in system entropy $\Delta s$, when the system begins and ends in a steady state: $\Delta\phi = -\ln[p_s(\bm x_\tau,\bm v_\tau;\tau)/p_s(\bm x_0,\bm v_0;0)]$.
 The above relation is the modified integral fluctuation theorem (MIFT) for the heat dissipated into the hot bath.

The other important relation is obtained as \cite{noh12,lah13a_arxiv}
\begin{align}
&\int d\Delta E~dW~d\Delta \phi ~P_f(\Delta E,W,\Delta \phi|Q_h) P_f(Q_h) \nn\\
&\hspace{3cm}\times~e^{-(\beta_h-\beta_c)Q_h+\beta_c(\Delta E-W)-\Delta \phi} \nn\\
&= \int d\Delta E~dW~d\Delta \phi ~P_r(-\Delta E,-W,-\Delta \phi|-Q_h) \nn\\
&\hspace{5cm}\times P_r(-Q_h)\nn\\
\end{align}
\normalsize
\begin{align}
&\Ra P_f(Q_h) \Psi(Q_h) e^{-(\beta_h-\beta_c)Q_h} = P_r(-Q_h) \nn\\
&\Ra 
\frac{P_f(Q_h)}{P_r(-Q_h)} = \frac{e^{(\beta_h-\beta_c)Q_h}}{\Psi(Q_h)},
\label{DFT_Qh}
\end{align}

where 
\begin{align}
\Psi(Q_h) &= \int d\Delta E~d\Delta \phi ~P_f(\Delta E,W,\Delta \phi|Q_h) \nn\\
&\hspace{2cm}~\times e^{\beta_c(\Delta E-W)-\Delta\phi}.
\end{align}

Eq. \eqref{DFT_Qh} provides the modified detailed fluctuation theorem (MDFT) for $Q_h$. The difference from the generic form of DFT is given by the factor $\Psi(Q_h)$. In case its value accidentally turns out to be equal to unity in some special case, we will have the usual DFT for this system. Also, it may be noted that for this model, the internal energy is unbounded. Even for a single heat bath, invalidity of steady state fluctuation theorem (SSFT) for large fluctuations of heat is due to the unbounded nature of the internal energy \cite{zon03}.

Although we have considered a one-dimensional system in our analysis, the generalization of the model to multiple dimensions is trivial.

We will now numerically investigate this model system for the simplistic case of two interacting particles, connected as usual to the two heat baths at different temperatures.

\section{Numerical results}

We now provide the results of our numerical simulations. For simplicity, instead of considering a chain of $n$ particles, we retain only two particles that are harmonically coupled. Here, we would like to mention that the harmonic coupling used in our simulations is only for convenience. The theorems proved in the previous sections are valid for any form of interaction between the particles.

The two coupled particles follow the following Langevin equations:
\begin{align}
m\dot{v_1} &= -\gamma_h v_1 -k_1(t) x_1 -K(x_1-x_2) +\xi_h(t); \nn\\
m\dot{v_2} &= -\gamma_c v_2 -k_2(t) x_2 -K(x_2-x_1) +\xi_c(t).
\end{align} 
Here, $\la\xi_h(t)\ra=\la\xi_c(t)\ra=0$, $\la \xi_h(t)\xi_h(t')\ra=2D_h\delta(t-t')$, and $\la \xi_c(t)\xi_c(t')\ra=2D_c\delta(t-t')$. $K$ is the interaction force between the two particles, while $k_1(t)$ and $k_2(t)$ are time-dependent spring constants of two harmonic traps in which the particles are placed. When the stiffness varies in time, so that work is done on the system. We will compare the two cases: (i) when $k_1(t)$ and $k_2(t)$ are constant in time (no work done), and (ii) when they are time-varying. Since we will analyze the systems only in their steady states, in case (i) we must have $\av{Q_h}=-\av{Q_c}$, since the entire heat absorbed from the hot bath is dissipated into the cold bath on average. On the other hand, in case (ii) work is done, so the magnitudes of the $\av{Q_h}$ and $\av{Q_c}$ will be different. 
It is to be noted that in principle even the IFT given by \eqref{IFT_Qh} can be verified. However, in this case it becomes too difficult, owing to the dependence of $\phi$ on four variables: $\phi=\phi(x_1,x_2,v_1,v_2;t)$.

In our simulations, we verify the above conditions, as well as the first law, given by $\av{Q_h+Q_c}=\av{W-\Delta E}$. The satisfaction of first law acts as a check on our simulations. The thermodynamic quantities are given by
\begin{align}
E(\tau) &= \frac{1}{2}\left[mv_1(\tau)^2+mv_2(\tau)^2+K(x_1(\tau)-x_2(\tau))^2\right.\nn\\
&\hspace{2cm}\left.+k_1(\tau)x_1^2(\tau)+k_2x_2^2(\tau)\right]; \nn\\
W(\tau) &= \frac{1}{2}\int_0^\tau dt \left[\dot k_1(t)x_1^2(t) + \dot k_2(t)x_2^2(t)\right] ; \nn\\
Q_h(t) &= \int_0^\tau dt ~v_1 [\gamma v_1-\xi_h(t)];\nn\\
Q_c(t) &= \int_0^\tau dt ~v_2 [\gamma v_2-\xi_c(t)].
\end{align}
To abide by the Stratonovich scheme, we need to take $v_1=[v_1(t)+v_1(t+\Delta t)]/2$, and similar convention for $v_2$.
The hot and cold baths are kept at temperatures $T_h=0.3$ and $T_c=0.1$, respectively. The friction coefficients of the two baths are considered to be equal to unity:  $\gamma_h=\gamma_c=1$. The time-dependent spring constants of the harmonic traps act as the external protocol. In case(i), we have considered the absence of a time-dependent protocol, and we have taken $k_1=k_2=1$. For case (ii), we choose the following time-dependence of the spring constants, with $k_1(t)=k_1(0)+\sin(\omega t)$ and $k_2(t)=k_2(0)+\cos(\omega t)$. We have taken $k_1(0)=k_2(0)=1.1$ and $\omega=1$, and our time of observation $\tau=3\tau_\omega$, where  $\tau_\omega$ is the time-period of this drive. The same time of observation has been used for case (i) as well (although there is no drive), in order to compare the results in the two cases. 

For our simulation purposes, we have used the Heun's method of integration \cite{man}, using $10^5$ realizations. 

\begin{figure}[!h]
\centering
\includegraphics[width=\lw]{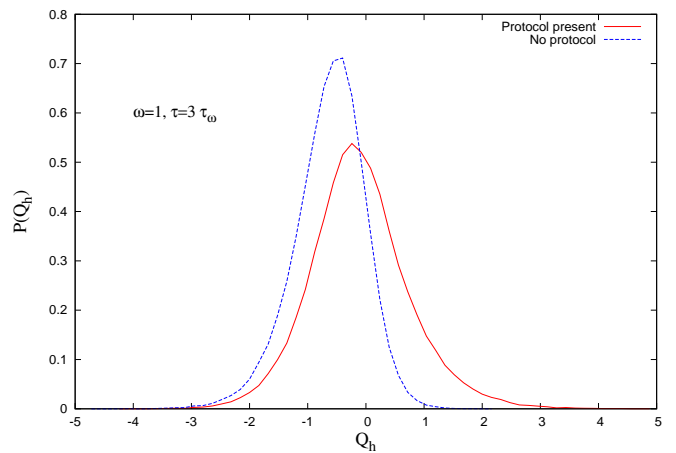}
\caption{Plot of probability distributions for $Q_h$, in absence as well as in presence of external protocol. In either case, the mean of the distributions is on the negative side (heat is absorbed from the hot bath). The parameters used are: $T_h=0.3$, $T_c=0.1$, $\gamma_h=\gamma_c=1$.}
\label{fig:P(Q_h)}
\end{figure}

\begin{figure}[!h]
\centering
\includegraphics[width=\lw]{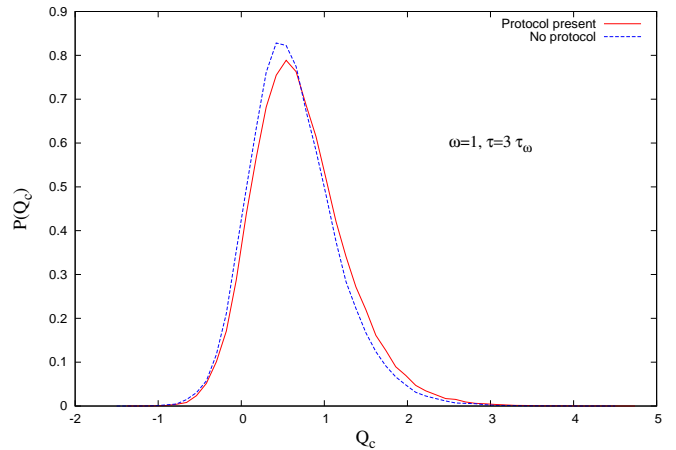}
\caption{Plot of probability distributions for $Q_c$, in absence as well as in presence of external protocol. In either case, the mean of the distributions is on the positive side (heat is absorbed from the hot bath),  for the same set of parameters as in figure \ref{fig:P(Q_h)}.}
\label{fig:P(Q_c)}
\end{figure}

As can be observed from figure \ref{fig:P(Q_h)}, the distributions of $Q_h$ for this system show a negative mean, which implies that on average, heat is absorbed from the hot bath. We find that the distributions $P(Q_h)$ have a finite weights in the positive side, which implies that during some realizations, heat is \emph{released} into the hot bath. Such realizations constitute the \emph{transient} violations to the second law \cite{lah11_jpa}. On the contrary, figure \ref{fig:P(Q_c)} shows that heat is released on average into the cold bath, but the distributions of $P(Q_c)$ have finite weights in the negative side. The distributions are non-Gaussian in all the cases.

Next, taking the logarithm of both sides of the DFT \eqref{DFT_Qh}, we get
\begin{align}
(\beta_h-\beta_c)Q_h = \mbox{Sym}(Q_h)+\ln\Psi(Q_h),
\end{align}
where
\[
\mbox{Sym}(Q_h) = \ln \frac{P(Q_h)}{P(-Q_h)}
\]
is the so-called symmetry function of $Q_h$. This implies that if we plot the symmetry function as a function of $Q_h$, we do not expect a straight line of slope 1, which is required for the validity of the detailed fluctuation theorem. This condition is only fulfilled if $\Psi(Q_h)$ becomes 1 for some system. In that case, the $Q_h$ must become independent of $\Delta E$, $W$ and $\Delta \phi$. This is an unlikely proposition, given the first law: $Q_h+Q_c = W-\Delta E$. 
\begin{figure}[!h]
\centering
\includegraphics[width=\lw]{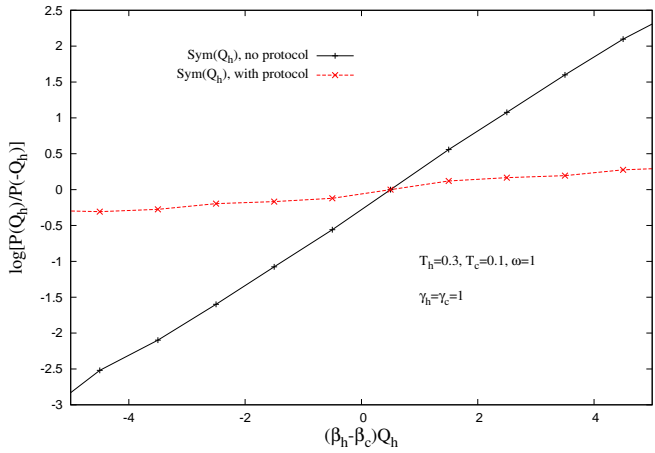}
\caption{Symmetry function of $Q_h$, both in presence as well as in absence of time-dependent protocols.}
\label{fig:Qhsym}
\end{figure}
\begin{figure}[!h]
\centering
\includegraphics[width=\lw]{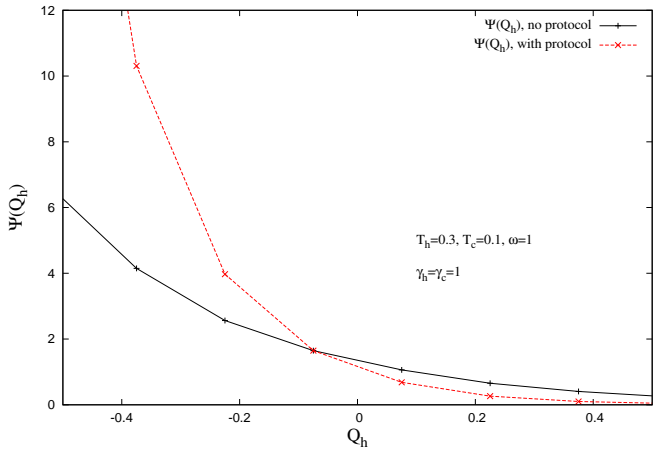}
\caption{Plot of $\Psi(Q_h)$ vs $Q_h$, both in presence as well as in absence of time-dependent protocol. This quantity measures the deviation from DFT of $Q_h$.}
\label{fig:PsiQh}
\end{figure}

In figure \ref{fig:Qhsym}, we have plotted $\mbox{Sym}(Q_h)$ as a function of $Q_h$. Clearly, the DFT is violated by $Q_h$. This is obvious, because of the extra factor $\Psi(Q_h)$ appearing in the MDFT, eq. \eqref{DFT_Qh}.

In figure \ref{fig:PsiQh}, we have plotted $\Psi(Q_h)$ as a function of $Q_h$, with the same set of parameters. We find that $\Psi(Q_h)$ is large when $Q_h$ is negative (heat is absorbed from hot bath), while it is small when $Q_h$ is negative (rare event: heat is dissipated to the hot bath).

Finally, we would like to state that after running our simulation for various parameter regimes, we could not find a set of parameters for which the system acts as a heat pump \cite{jay07_pre} (where heat flows from the cold to the hot reservoir).

\section{Conclusions}

In this work, we have derived the fluctuation theorems for heat exchanged with each bath, for a chain of interacting particles connected at each end to Langevin baths, in presence of an external drive. The heat absorbed by the first particle from the hot bath is $Q_h$, and by the last particle from the cold bath is $Q_c$. Both these variables follow the modified detailed fluctuation theorems, that differs from a true DFT due to the presence of the factor $\Psi(Q)$ in the denominator (see eq. \eqref{DFT_Qh}). This is a generalization of the exchange fluctuation theorem obtained when the two subsystems are in direct contact with each other, as in \cite{jar04}. We have numerically studied this model, with only two particles with harmonic interaction. The symmetry functions for $Q_h$ obtained from our simulations show that the DFT is violated to a higher extent, when an external perturbation acts on the particles. 
In this case, it may be noted that the baths exchange different amounts of heat.

\section{Acknowledgement}

One of us (AMJ) thanks DST, India for financial support.



\end{document}